\documentclass[%
reprint,
superscriptaddress,
showpacs,
 amsmath,amssymb,
 aps,
 pre,
]{revtex4-1}
\pdfoutput=1
\usepackage[T1]{fontenc}
\usepackage[utf8]{inputenc}
\usepackage{lmodern}
\usepackage{textcomp}
\usepackage{todonotes}
\usepackage{graphicx}
\usepackage{dcolumn}
\usepackage{bm}
\usepackage{hyperref}
\hypersetup{%
  colorlinks=true,
  citecolor=blue,%
  urlcolor=blue,
  linkcolor=blue
}
\usepackage[capitalise]{cleveref}
\usepackage{mathtools}

\usepackage{tikz}
\usepackage{tikzscale}
\usetikzlibrary{chains}
\usetikzlibrary{positioning}

\usepackage{etoolbox}
\makeatletter
\appto{\appendix}{%
  \@ifstar{\def\theequation@prefix{A.}}%
          {}%
}
\makeatother

\newcommand*\mean[1]{\bar{#1}}
\newcommand*\enmean[1]{\langle{#1}\rangle}
\begin{document}


\title{Relaxation dynamics of maximally clustered networks}

\author{Janis Klaise}
 \email{J.Klaise@warwick.ac.uk}
 \affiliation{Centre for Complexity Science, University of Warwick, Coventry CV4 7AL, United Kingdom}
\author{Samuel Johnson}%
 \email{S.Johnson.2@warwick.ac.uk}
\affiliation{Warwick Mathematics Institute, University of Warwick, Coventry CV4 7AL, United Kingdom}%

\date{\today}

\begin{abstract}
We study the relaxation dynamics of fully clustered networks (maximal number of triangles) to an unclustered state under two different edge dynamics---the double-edge swap, corresponding to degree-preserving randomization of the configuration model, and single edge replacement, corresponding to full randomization of the Erd\H{o}s--R\'enyi random graph. We derive expressions for the time evolution of the degree distribution, edge multiplicity distribution and clustering coefficient. We show that under both dynamics networks undergo a continuous phase transition in which a giant connected component is formed. We calculate the position of the phase transition analytically using the Erd\H{o}s--R\'enyi phenomenology.
\end{abstract}

\maketitle


\section{\label{sec:introduction}Introduction}
Network science has enjoyed an unprecedented popularity in the last two decades as a paradigm for studying complex, interacting systems such as the Internet~\citep{yook2002,johnson_entropic_2010}, World Wide Web~\citep{barabasi_emergence_1999}, food webs~\citep{johnson_trophic_2014}, scientific collaboration networks~\citep{newman_scicol_2001}, social and biological networks~\citep{robins_introduction_2007,albert2000large,bullmore2009complex}, contact networks~\citep{danon_social_2012,pastor-satorras_epidemic_2015} and many others~\citep{boccaletti_complex_2006}. Many of these empirical networks exhibit a high degree of \emph{clustering} or \emph{transitivity}, i.e. a significant number of short, closed loops forming triangles~\citep{wasserman1994social,watts1998collective}. This phenomenon is most commonly quantified by the clustering coefficient, defined as the proportion of connected triads that are also triangles in a network~\citep{newman_introduction_2010}.

The classical random network models, the Erd\H{o}s--R\'enyi random graph~\citep{gilbert_1959,erdos1959random} and the configuration model~\citep{newman_introduction_2010,molloy_reed_1995,bollobas_2001}, both suffer from being unable to generate networks with significant values of the clustering coefficient thus making them unsuitable for modelling many real networks. High values of the clustering coefficient observed in empirical networks have lead to a surge of random network models that are capable of generating significant numbers of triangles~\citep{watts1998collective,newman_properties_2003,newman_random_2009,volz_random_2004,park_solution_2005}. The relationship between clustering and other network properties has also been studied extensively~\citep{del_genio_endemic_2013,serrano_clustering_2006,serrano_clustering_2006-1,serrano_percolation_2006,vazquez_topological_2004,foster_communities_2010,berchenko_emergence_2009,foster_clustering_2011}. However, despite the large body of research, the inherent violation of edge independence in highly clustered networks has made it difficult to understand the full implications of clustering. Common issues encountered when dealing with highly clustered networks include difficulties of network sampling~\citep{park_solution_2005,horvat_reducing_2015}, inability to use edge independence to derive accurate results~\citep{serrano_percolation_2006,berchenko_emergence_2009} and potentially overstated inferences of causality~\citep{foster_clustering_2011,foster_communities_2010}. This points to a need for more fundamental research in clustered networks.

In this paper we explore simple dynamics of highly clustered networks relaxing to an unclustered equilibrium state. Specificaly, we study the evolution of the clustering coefficient under two edge rewiring schemes starting with fully clustered, degree-regular networks, i.e. networks in which all nodes have the same number of neighbours and a maximal number of triangles. We find that under both dynamics whose equilibrium distributions correspond to the Erd\H{o}s--R\'enyi random graph and the configuration model respectively, a giant connected component emerges via a continuous phase transition. We provide an analytical prediction of the critical point for this transition as well as derive time evolution equations for various network properties.

\section{\label{sec:methods}Methods}
\subsection{Network metrics}
We consider undirected graphs with $N$ nodes and $L$ edges described by a symmetric $N\times N$ adjacency matrix $\mathbf{A}$ with binary edge variables $A_{ij}\in\lbrace 0,1\rbrace$ for $i,j\in\lbrace 1,\dots,N\rbrace$ with $A_{ij}=1, i\neq j$ indicating an edge between nodes $i$ and $j$ so that $L=\sum_{i,j}A_{ij}$. The degree distribution of a network is defined as $p_k=N_k/N$, where $N_k$ is the number of nodes with degree $k$. We denote the $n$th moment of the degree distribution by $\enmean{k^n}$.

We define the \emph{multiplicity} $m_{ij}$ of an edge $ij$ to be the number of triangles it participates in~\citep{serrano_clustering_2006}. Similarly to the degree distribution, we define the edge multiplicity (or simply multiplicity) distribution as $q_m=L_m/L$, where $L_m$ is the number of edges with multiplicity $m$. We denote the $n$th moment of the multiplicity distribution by $\enmean{m^n}$.

The clustering coefficient of a network is defined as three times the number of triangles divided by the number of connected triples, i.e. $C=3N_{\triangle}/N_{\wedge}$~\citep{newman_introduction_2010}. This measure of clustering is properly normalized so that $C\in[0,1]$. It also admits a probabilistic interpretation---it is the probability that a randomly chosen triple of nodes is closed.

We can express the clustering coefficient in terms of the degree and multiplicity distributions. For any network we have
\begin{equation}
  N_{\wedge} = \sum_{k}\binom{k}{2}N_k = N\sum_{k}\binom{k}{2}p_k = N\frac{\enmean{k^2}-\enmean{k}}{2}
\end{equation}
and
\begin{equation}
  3N_{\triangle} = \sum_{m}L_m = L\sum_{m}mq_m = L\enmean{m}.
\end{equation}
Putting the above results together and noting that in any network $L=N\enmean{k}/2$, we obtain the following general expression for the clustering coefficient:
\begin{equation}\label{eq:clustering}
  C = \frac{\enmean{k}\enmean{m}}{\enmean{k^2}-\enmean{k}}.
\end{equation}

\subsection{Random network ensembles}
We study relaxation dynamics of $k$-regular networks under edge rewiring in two random network ensembles---the configuration model (CM) and the Erd\H{o}s--R\'enyi random graph (ER).

The CM~\citep{molloy_reed_1995,bollobas_2001,newman_introduction_2010} is defined by drawing a valid degree sequence $\mathbf{k}=\lbrace k_i\rbrace_{i=1}^{N}$ from a degree distribution $p_k$ and producing a network realization uniformly at random from all possible networks with that degree sequence~\citep{newman_introduction_2010,del2010efficient}. Provided the second moment of the degree distribution remains finite, it can be shown that the clustering coefficient scales as $C\sim 1/N$ so that in the thermodynamic limit ($N\to\infty$) the resulting networks are tree-like~\citep{newman_introduction_2010}.

The ER random graph~\citep{gilbert_1959,erdos1959random,erdos1960evolution} is defined by placing $L$ edges uniformly at random between $N$ nodes~\footnote{Another common definition leading to a slightly different model is to place each of the possible $\binom{N}{2}$ edges with equal probability $p$, but this does not enforce a fixed number of edges.}. If we require that the mean degree $\enmean{k}=2L/N$ be fixed, the degree distribution of the ER model in the thermodynamic limit is Poisson with mean $\enmean{k}$~\citep{newman_introduction_2010}. The ER model is thus a special case of the CM and has the same scaling behaviour of the clustering coefficient.

\begin{figure}[htbp]
  \centering
  \includegraphics[width=\linewidth]{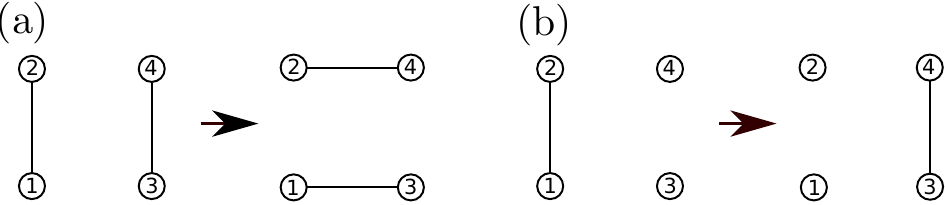}
  \caption{(a) Double edge swap or degree-preserving randomization. (b) Edge replacement or full randomization.}
  \label{fig:rewiring}
\end{figure}

Given that both the CM and ER random graphs are asymptotically triangle-free, it is natural to consider them as equilibrium ensembles for relaxation dynamics of highly clustered networks into an unclustered state. To this end we describe two edge rewiring mechanisms that have the CM and the ER random graphs as equilibrium distributions (see \cref{fig:rewiring} for a graphical demonstration).

\paragraph*{Double edge swap (CM).}
The double-edge swap~\citep{ramachandra_1996,milo_uniform_2003} is defined by choosing two existing edges in the network at random and rewiring their ends to produce two new edges while deleting the original two. This is also known as \emph{degree-preserving} randomization and so naturally produces network realizations in the CM ensemble with a fixed degree sequence. The double-edge swap defines a Markov chain whose equilibrium distribution is the CM~\citep{ramachandra_1996}.

\paragraph*{Edge replacement (ER).}
Alternatively, one can fully randomize a network by picking an edge at random and placing it anywhere in the network where there is no edge already~\citep{liu2011controllability,Nepusz2012}. In this scheme the number of edges is preserved but the degrees of the nodes are not. Edge replacement defines a Markov chain whose equilibrium distribution is the ER ensemble.

A double-edge swap or an edge replacement constitutes an \emph{elementary rewiring step}.

\section{\label{sec:results}Results}
To assess the evolution of network measures over time, we take into account the network size and the rewiring scheme (either CM or ER) to normalize the number of elementary rewiring steps per number of edges. If $r_{\text{CM}}$ and $r_{\text{ER}}$ are the number of elementary rewiring steps in the CM and ER ensembles respectively, we define the corresponding time variables as
\begin{equation}\label{eq:times}
  \begin{aligned}
    t_{\text{ER}} &= \frac{r_{\text{ER}}}{L}\\
    t_{\text{CM}} &= \frac{2r_{\text{CM}}}{L}.
  \end{aligned}
\end{equation}
These definitions have the useful interpretation that when $t_{\text{scheme}}=1$, the rewiring scheme has, on average, modified each edge in the network.

\subsection{Multiplicity distribution}
The multiplicity distribution evolves over time as edges are rewired and triangles are destroyed. The initial configuration of a $k$-regular network is a disjoint union of $N/(\enmean{k}+1)$ cliques of size $\enmean{k}+1$ which ensures maximal clustering $C=1$. In other words, at time $t=0$, the multiplicity distribution is
\begin{equation}
  \begin{dcases}
    q_{\enmean{k}-1} &= 1 \\
    q_m &=0 \text{ if $m\neq \enmean{k}-1$}.
  \end{dcases}
\end{equation}

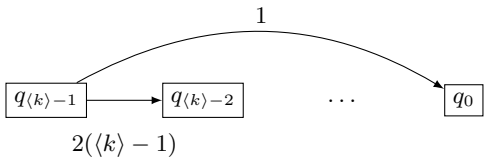
\begin{figure}
\centering
  \begin{tikzpicture}[
  mynode/.style={
    draw,
    minimum size=1em
    },
  every loop/.append style={-latex},
  start chain=going right
  ]
  \node[mynode,on chain] (qk1) {$q_{\enmean{k}-1}$};
  \node[mynode,on chain] (qk2) {$q_{\enmean{k}-2}$};
  \node[on chain] (dots) {$\dots$};
  \node[mynode,on chain] (q0) {$q_{0}$};

  \path[-latex]
    (qk1) edge[] node[below =1em of qk1,font=\small] {$2(\enmean{k}-1)$} (qk2)
    (qk1) edge[bend left] node[auto,font=\small] {$1$} (q0);r
  \end{tikzpicture}
  \caption{Transition rates in the multiplicity distribution for a single clique of size $\enmean{k}+1$.} \label{fig:mult_clique}
\end{figure}

Consider the smallest informative time step $\Delta t_{\text{CM}}=2/L$ or $\Delta t_{\text{ER}}=1/L$ corresponding to exactly one elementary rewiring step. At $t=0$ a clique of size $\enmean{k}+1$ has exactly $\binom{\enmean{k}+1}{2}$ edges all of which have maximal multiplicity $\enmean{k}-1$. Rewiring any single edge will destroy $\enmean{k}-1$ triangles leading to a decrease of $2(\enmean{k}-1)+1$ edges with maximal multiplicity, one for the rewired edge and an additional two for each destroyed triangle. Assuming that no new triangles are created, the single rewired edge will have multiplicity zero. \Cref{fig:mult_clique} shows the transition rates in the multiplicity distribution of a single clique.

\begin{figure}
\centering
\resizebox{\linewidth}{!}{%
  \begin{tikzpicture}[
  mynode/.style={
    draw,
    minimum size=1em
    },
  every loop/.append style={-latex},
  start chain=going right,
  ]
  \node[mynode,on chain] (qk1) {$q_{\enmean{k}-1}$};
  \node[mynode,on chain] (qk2) {$q_{\enmean{k}-2}$};
  \node[on chain] (dots) {$\dots$};
  \node[mynode,on chain] (q2) {$q_{2}$};
  \node[mynode,on chain] (q1) {$q_{1}$};
  \node[mynode,on chain] (q0) {$q_{0}$};

  \path[-latex]
    (qk1) edge[] node[below = 1em of qk1,swap] {$2(\enmean{k}-1)$} (qk2)
    (qk2) edge[] node[above = 0.75em of qk2,swap] {$2(\enmean{k}-2)$} (dots)
    (dots) edge[] node[auto,swap] {$6$} (q2)
    (q2) edge[] node[auto,swap] {$4$} (q1)
    (q1) edge[] node[auto,swap] {$3$} (q0)
    (qk1) edge[bend left=40] node[auto,swap] {$1$} (q0)
    (qk2) edge[bend right=35] node[auto,swap] {$1$} (q0)
    (q2) edge[bend right=30] node[auto,swap] {$1$} (q0);
  \end{tikzpicture}
}
  \caption{Transition rates in the full multiplicity distribution.} \label{fig:mult_full}
\end{figure}
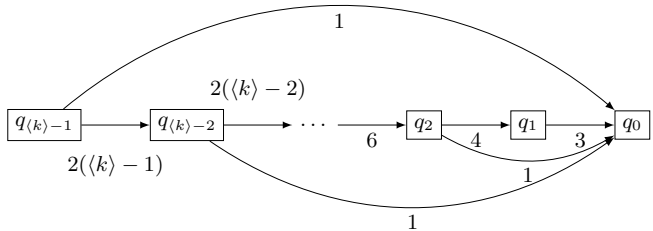
We now make the \emph{ansatz} that this is the main way the multiplicity distribution changes over time---multiplicity is predominantly decreased by rewiring single edges from cliques and all such rewirings are independent. In this case, we can write down the full transition rate diagram between multiplicity classes as shown in \cref{fig:mult_full}. This gives the following time evolution equations for $q_m$:
\begin{equation}\label{eq:dt_mult}
  \begin{dcases}
    \frac{dq_{m}}{dt} &= -(2m+1)q_m+2(m+1)q_{m+1}, \begin{array}{@{}l@{}}\text{ for $m=$}\ \\\,\enmean{k}-1,\dots, 1\end{array}\\
    \frac{dq_{0}}{dt} &= 3q_1+\sum_{m=2}^{\enmean{k}-1}q_m.
  \end{dcases}
\end{equation}

\begin{figure}[tb]
  \centering
  \includegraphics[]{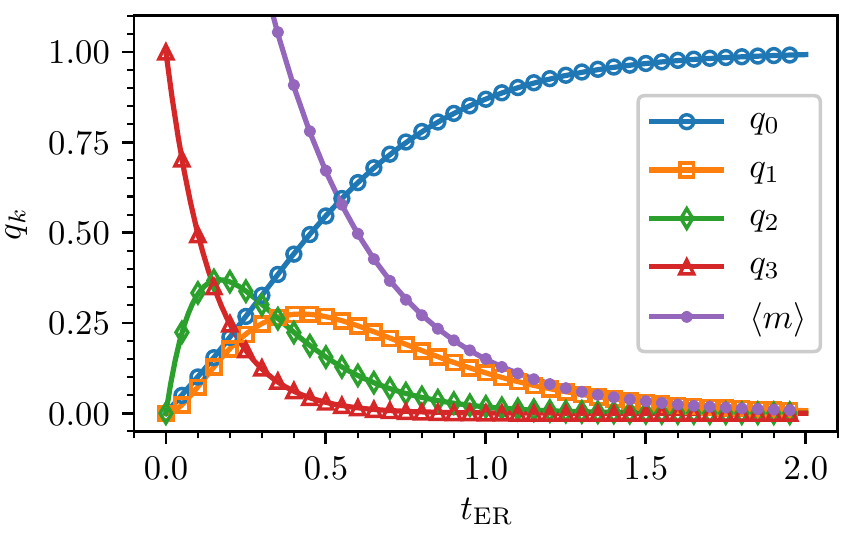}
  \caption{Evolution of the multiplicity distribution in an ER network with average degree $\enmean{k}=4$ and $N=10^5$. The solid lines are numerical solutions of \cref{eq:dt_mult} while the markers are simulation results. The purple line with filled circles indicates the average multiplicity $\enmean{m}$.}
  \label{fig:4_reg_mul}
\end{figure}

\Cref{fig:4_reg_mul} shows the numerical solution of these ODEs which is in excellent agreement with simulation results. The calculations are valid both in the ER and the CM case.

\paragraph*{Average multiplicity.}
Using the time evolution equations for the multiplicity distribution, we can derive exact expressions of its moments. Specifically we are interested in the average multiplicity $\enmean{m}$ as it features in the expression for the clustering coefficient. We have
\begin{equation}
  \frac{d\enmean{m}}{dt} = \sum_{m=1}^{\enmean{k}-1}m\frac{dq_m}{dt}.
\end{equation}
Inserting \cref{eq:dt_mult} we obtain the simple expression
\begin{equation}
  \frac{d\enmean{m}}{dt} = -3\enmean{m}.
\end{equation}
Using the initial condition $\enmean{m}(0)=\enmean{k}-1$, this has solution
\begin{equation}
  \enmean{m} = (\enmean{k}-1)e^{-3t}.
\end{equation}
\Cref{fig:mult_full} shows the analytic solution of the average multiplicity which is in perfect agreement with simulation results.

\subsection{Degree distribution}
In the case of the ER model, the degree distribution is also changing over time. Consider the degree distribution $p_k(t)$ as a function of time and a time step $\Delta t_{\text{ER}}$. We can calculate the rate at which $p_k(t)$ changes.

An edge replacement event in the ER model consists of two steps. First, a random edge is selected. Second, a random pair of nodes that are not linked by an edge (let us call this pair a \emph{non-edge}) is selected and the edge selected in the first step is deleted while the non-edge becomes an edge.

When a random edge is selected, $p_k$ can decrease if at least one end of the edge has degree $k$. Alternatively, $p_k$ can increase if at least one end of the edge has degree $k+1$. The probability of reaching a node of degree $k$ by following a randomly chosen edge is given by the so called excess degree distribution~\citep{newman_introduction_2010} which reads $s_k=kp_k/\enmean{k}$. Given this and the fact that a randomly chosen edge can have 0,1 or 2 nodes of degree $k$, we can calculate the expected number of nodes of degree $k$ at the ends of a random edge:
\begin{equation}
  \mathbb{E}\left(k\rightarrow k-1\right)=2s_k^2+2s_k\left(1-s_k\right)=2s_k=2\frac{kp_k}{\enmean{k}}.
\end{equation}
This is the expected number of nodes whose degree would decrease from $k$ to $k-1$ during a single edge selection step. Note that at the beginning of the process the degree distribution is regular so $\mathbb{E}\left(k\rightarrow k-1\right)=2$ as expected.

Similarly, the expected number of nodes whose degree would decrease from $k+1$ to $k$ leading to an increase in $p_k$ is:
\begin{equation}
  \mathbb{E}\left(k+1\rightarrow k\right)=2s_{k+1}=2\frac{\left(k+1\right)p_{k+1}}{\enmean{k}}.
\end{equation}

Now consider the second step in the edge replacement event, the selection of a non-edge. When a random non-edge is selected, $p_k$ can also change in two ways. It can increase if at least one of the selected nodes has degree $k-1$ and it can decrease if at least one of the nodes has degree $k$. The calculation of the expected number of nodes changed as a result of this is similar to the previous case, but we must consider the distribution of \emph{non-degrees} instead. To this end we study the graph complement of the original network defined as a network in which two nodes are linked if and only if they are not linked in the original network. From here on we denote by an overbar quantities in the graph complement.

It is easy to see that the degrees of nodes in the complement are given by $\mean{k}=N-1-k$ where $k$ is the degree of a node in the original network and we have $p_{\mean{k}}=p_k$. Thus, the non-edges are selected proportionally to $\mean{k}$ not $k$ as in the case of edge selection so we must work with the excess non-degree distribution given by $s_{\mean{k}}=\mean{k}p_k/\enmean{\mean{k}}$. Note that the mean non-degree is given by
\begin{equation}
  \enmean{\mean{k}} = \sum_k\mean{k}p_k = N-1-\enmean{k}.
\end{equation}
As in the case of edge selection, the expected number of nodes whose degree would increase from $k$ to $k+1$ thus reducing $p_k$ during a single non-edge selection step is
\begin{equation}
  \mathbb{E}\left(k\rightarrow k+1\right)=2q_{\mean{k}}=2\frac{\mean{k}p_k}{\enmean{\mean{k}}} = \frac{2\left(N-1-k\right)}{N-1-\enmean{k}}p_k.
\end{equation}
When $N$ is large we can approximate this by
\begin{equation}
  \mathbb{E}\left(k\rightarrow k+1\right)\simeq 2p_k.
\end{equation}

Similarly, $p_k$ can increase if we select a non-edge with at least one node with degree $k-1$. The expected number of such nodes in a single non-edge selection is
\begin{equation}
  \mathbb{E}\left(k-1\rightarrow k\right)\simeq 2p_{k-1}.
\end{equation}

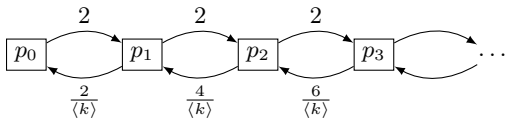
\begin{figure}
\centering
  \begin{tikzpicture}[
  mynode/.style={
    draw,
    minimum size=1em
    },
  every loop/.append style={-latex},
  start chain=going right
  ]
  \node[mynode,on chain] (p0) {$p_{0}$};
  \node[mynode,on chain] (p1) {$p_{1}$};
  \node[mynode,on chain] (p2) {$p_{2}$};
  \node[mynode,on chain] (p3) {$p_{3}$};
  \node[on chain] (dots) {$\dots$};

  \path[-latex]
    (p0) edge[bend left] node[auto] {$2$} (p1)
    (p1) edge[bend left] node[auto] {$\frac{2}{\enmean{k}}$} (p0)
    (p1) edge[bend left] node[auto]  {$2$} (p2)
    (p2) edge[bend left] node[auto] {$\frac{4}{\enmean{k}}$} (p1)
    (p2) edge[bend left] node[auto] {$2$} (p3)
    (p3) edge[bend left] node[auto] {$\frac{6}{\enmean{k}}$} (p2)
    (p3) edge[bend left] node[auto] {} (dots)
    (dots) edge[bend left] node[auto] {} (p3);
  \end{tikzpicture}
  \caption{Transition rates in the degree distribution under the ER model.} \label{fig:deg_rates}
\end{figure}

\begin{figure}[htbp]
  \centering
  \includegraphics[]{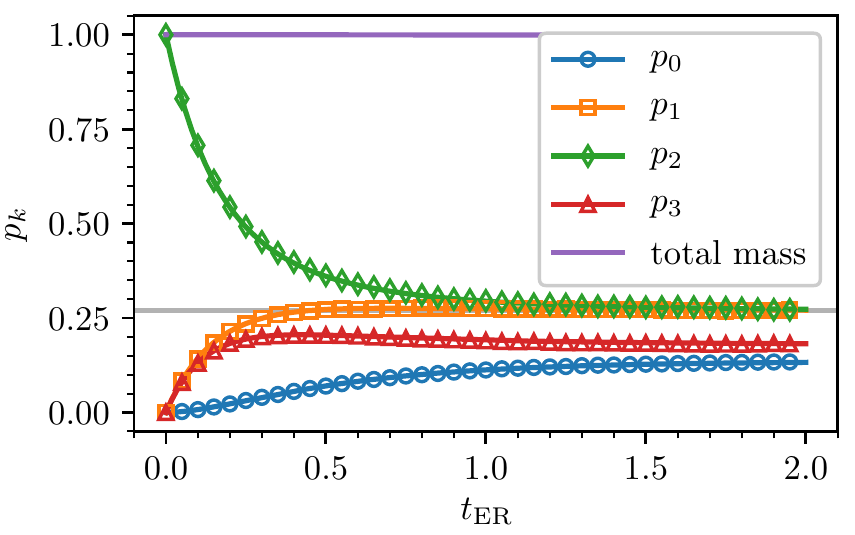}
  \caption{Evolution of the degree distribution in an ER network with average degree $\enmean{k}=2$ and $N=10^5-1$. The solid lines are numerical solutions of \cref{eq:dt_deg} while the markers are simulation results. The grey line indicates the equilibrium value of $p_2$ in an ER ensemble. The purple line indicates the total probability mass in the system accounted for by truncating the ODE system at $k^*=8$.}
  \label{fig:2_reg_deg}
\end{figure}

\Cref{fig:deg_rates} describes pictorially the transition rates between degree classes as derived here. This allows us to write down the time evolution equations for $p_k$:
\begin{equation}\label{eq:dt_deg}
  \frac{dp_k}{dt} = 2p_{k-1}-2\left(1+\frac{k}{\enmean{k}}\right)p_k+2\frac{k+1}{\enmean{k}}p_{k+1},
\end{equation}
for $k=0,1,\dots$. This system of ODEs is not closed, so in order to solve it numerically, we must truncate the system at some $p_{k^*}$ setting $p_k=0$ for all $k>k^*$. The value of $k^*$ should be set high enough so the probability mass unaccounted for is minimal for accurate predictions. We test our predictions by numerically solving the ODEs for a network with average degree $\enmean{k}=2$ and setting the cut-off $k^*=8$. The results are shown in \cref{fig:2_reg_deg}. The numerical solution of the ODE system is in excellent agreement with simulation results. We also note that the cut-off is appropriate for this level of approximation as the total probability mass does not diverge from unity noticeably over the time period considered.

\paragraph*{Second moment of the degree distribution.}
Using the time evolution equations for the degree distribution, we can derive exact expressions of its moments. Specifically, we are interested in the second moment $\enmean{k^2}$. We have
\begin{equation}
  \frac{d\enmean{k^2}}{dt} = \sum_{k}k^2\frac{dp_k}{dt}.
\end{equation}
Inserting \cref{eq:dt_deg} we obtain the simple expression
\begin{equation}
  \frac{d\enmean{k^2}}{dt} = -4\frac{\enmean{k^2}}{\enmean{k}}+4\enmean{k}+4.
\end{equation}
Using the initial condition $\enmean{k^2}(0) = \enmean{k}^2$ and recalling that the average degree $\enmean{k}$ is constant, this has solution
\begin{equation}
  \enmean{k^2} = \enmean{k}\left(\enmean{k}+1-e^{-\frac{4t}{\enmean{k}}}\right).
\end{equation}

\subsection{Clustering coefficient}
Putting together the results for the multiplicity and degree distributions, and using \cref{eq:clustering}, we obtain exact expressions for the clustering coefficient as a function of time in both the CM and ER ensembles:
\begin{equation}
  \begin{aligned}
    C_{\text{CM}} &= e^{-3t} \\
    C_{\text{ER}} &= \frac{\left(\enmean{k}-1\right)e^{-3t}}{\enmean{k}-e^{\frac{-4t}{\enmean{k}}}}.
  \end{aligned}
\end{equation}

We note that in the CM ensemble, the clustering coefficient has no dependence on the average degree while this is not the case for the ER ensemble. This is because the number of connected triples $N_{\wedge}$ in the CM ensemble is constant by virtue of having a fixed degree sequence while it is dependent on the evolving degree sequence in the ER ensemble.

\subsection{Giant connected component}
\begin{figure}[htbp]
  \centering
  \includegraphics[]{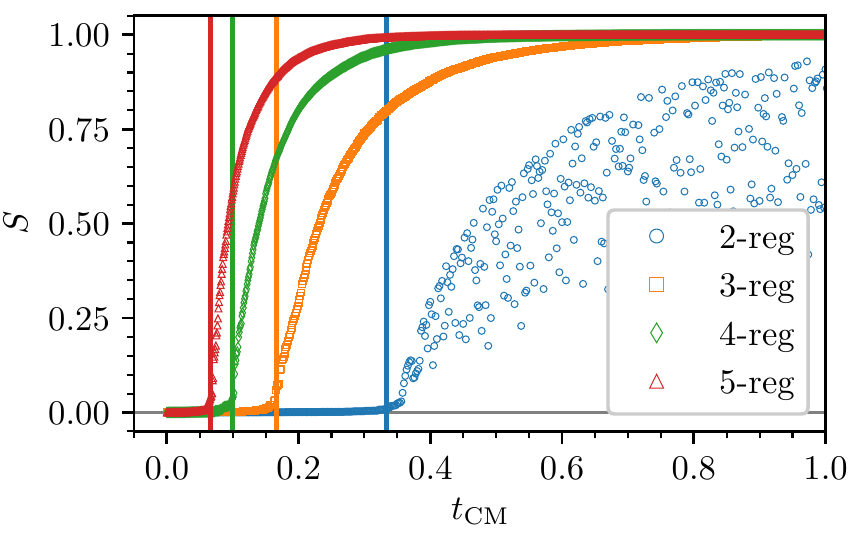}
  \caption{Proportion of nodes  $S$ in the giant connected component as a function of time $t_{\text{CM}}$ for a few select $k$-regular networks. We observe a continuous phase transition at a critical point $t^{\text{c}}_{\text{CM}}$ which depends on the average degree of the network as explained in the main text. Vertical lines correspond to the analytically calculated critical points.}
  \label{fig:s_vs_t}
\end{figure}

\begin{figure}[htbp]
  \centering
  \includegraphics[]{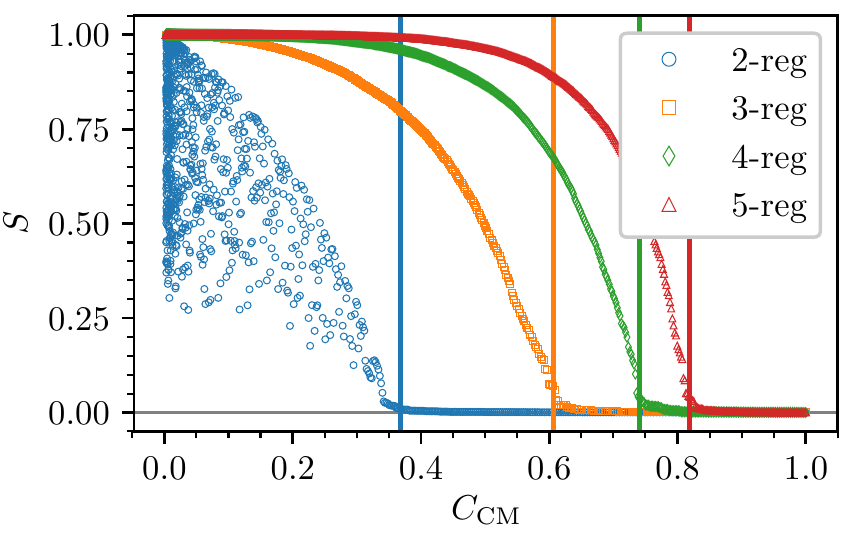}
  \caption{Proportion of nodes $S$ in the giant connected component as a function of clustering $C_{\text{CM}}$ for a few select $k$-regular networks under the CM rewiring scheme. We observe a continuous phase transition at a critical point $C^{\text{c}}_{\text{CM}}$ which depends on the average degree of the network as explained in the main text. Vertical lines correspond to the analytically calculated critical points.}
  \label{fig:s_vs_c}
\end{figure}

\begin{figure}[htbp]
  \centering
  \includegraphics[]{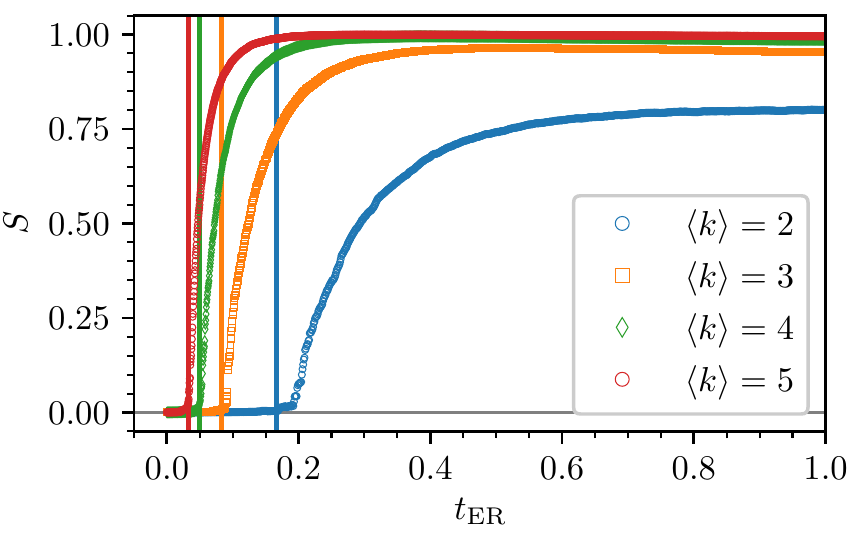}
  \caption{Proportion of nodes $S$ in the giant connected component as a function of time $t_{\text{ER}}$ for a few select mean degree $\enmean{k}$ networks. We observe a continuous phase transition at a critical point $t^{\text{c}}_{\text{ER}}$ which depends on the average degree of the network as explained in the main text. Vertical lines correspond to the analytically calculated critical points.}
  \label{fig:s_vs_t_er}
\end{figure}

\begin{figure}[htbp]
  \centering
  \includegraphics[]{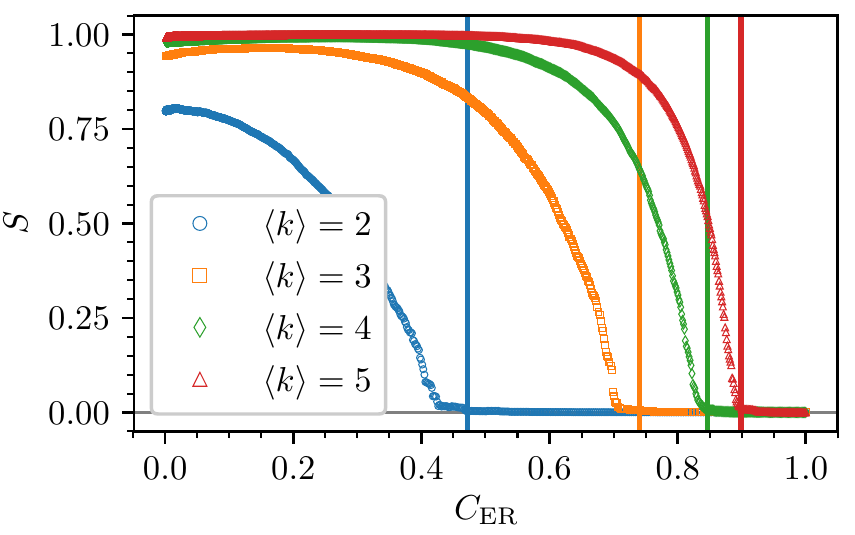}
  \caption{Proportion of nodes $S$ in the giant connected component as a function of clustering $C_{\text{ER}}$ for a few select mean degree $\enmean{k}$ networks under the ER rewiring scheme. We observe a continuous phase transition at a critical point $C^{\text{c}}_{\text{ER}}$ which depends on the average degree of the network as explained in the main text. Vertical lines correspond to the analytically calculated critical points.}
  \label{fig:s_vs_c_er}
\end{figure}

We find that under both rewiring schemes there is an emergence of global connectivity via the appearance of a giant connected component (GCC) at some critical time $t^c$ (equivalently, critical clustering coefficient $C^c$). We confirm from simulation results that a GCC emerges in a continuous phase transition (\cref{fig:s_vs_t,fig:s_vs_c} for the CM and \cref{fig:s_vs_t_er,fig:s_vs_c_er} for the ER ensembles). Note that the large fluctuations in the 2-regular case is due to the fact that 2-regular networks are exactly at the poing of criticality in the unclustered CM case ($C=0$). This phenomenon has been studied in the context of reversible polymerization of rings~\citep{ben-naim_kinetics_2011}.

We can calculate the critical point analytically by using the known result that a GCC in an ER random graph emerges when $\enmean{k}=1$~\citep{newman_introduction_2010}. We conjecture that a GCC induced by edge rewiring emerges when the average number of external edges between the original $N/(\enmean{k}+1)$ cliques of size $\enmean{k}+1$ exceeds one. If this is the case, the critical number of elementary rewiring steps is
\begin{equation}
  r^{\text{c}} = \frac{N}{2\left(\enmean{k}+1\right)}.
\end{equation}
Expressing this in terms of the time variable, we obtain the critical time for both the CM and the ER rewiring schemes:
\begin{equation}
  \begin{aligned}
    t^\text{c}_{\text{CM}} &= \frac{2}{\enmean{k}\left(\enmean{k}+1\right)} \\
    t^\text{c}_{\text{ER}} &= \frac{1}{\enmean{k}\left(\enmean{k}+1\right)}.
  \end{aligned}
\end{equation}
Note that these differ by a factor of two. This is because in the CM rewiring scheme, even though every elementary rewiring step involves two edges, the two rewirings are not independent---during one rewiring step it is possible to connect at most two disconnected components.

Expressed in terms of the clustering coefficient, the critical thresholds read:
\begin{equation}
  \begin{aligned}
    C^\text{c}_{\text{CM}} &= e^{-6/\enmean{k}\left(\enmean{k}+1\right)} \\
    C^\text{c}_{\text{ER}} &= \frac{(\enmean{k}-1)e^{-3/\enmean{k}(\enmean{k}+1)}}{\enmean{k}-e^{-4/\enmean{k}^2(\enmean{k}+1)}}.
  \end{aligned}
\end{equation}

\cref{fig:s_vs_t,fig:s_vs_c} confirm that these are in excellent agreement with simulations in the CM case and \cref{fig:s_vs_t_er,fig:s_vs_c_er} confirm a good agreement in the ER case which improves as the mean degree increases.

What is the cause of the discrepancy of the analytical result for the critical point and the numerical simulations, particularly for low mean degree ER networks? We conjecture that this is due to some edges being rewired multiple times while others are not rewired at all. This would have the effect of increasing the critical time because we have to wait slightly longer until the average number of rewired edges \emph{discouting edges rewired multiple times} reaches the point where long range connectedness emerges. \Cref{fig:s_vs_t_er} seems to confirm this to be the case. Let us calculate this revised critical time in the ER case.

During an edge replacement step, the probability of any edge being chosen for rewiring is $1/L$. So after $r$ rewiring events the probability that a specific edge has not been rewired is
\begin{equation}
  \mathbb{P}\left(\text{not rewired}\right) = \left(1-\frac{1}{L}\right)^r.
\end{equation}
Substituting $r=Lt$ since we are in the ER case and taking the limit as $L\to\infty$, we get
\begin{equation}
  \mathbb{P}\left(\text{not rewired}\right) = e^{-t}.
\end{equation}
The new revised time for ethe mergence of the GC, call it $t^{\text{r}}$, is then the time at which point this probability drops below a certain threshold. What is this threshold? It should be when the proportion of edges that have been rewired gives rise to a GCC which is precisely given by $t^{\text{c}}$. We can then write
\begin{equation}
  e^{-t^{\text{r}}} = 1-t^{\text{c}}.
\end{equation}
Note that by Taylor expansion we have $t^{\text{r}}\simeq t^{\text{c}}$ if this time is small as in the case when the average degree $\enmean{k}\to\infty$. This explains why the $t^{\text{c}}$ value becomes a better predictor for the critical threshold as the mean degree increases as seen in \cref{fig:s_vs_t_er}.

\begin{figure}[htbp]
  \centering
  \includegraphics[]{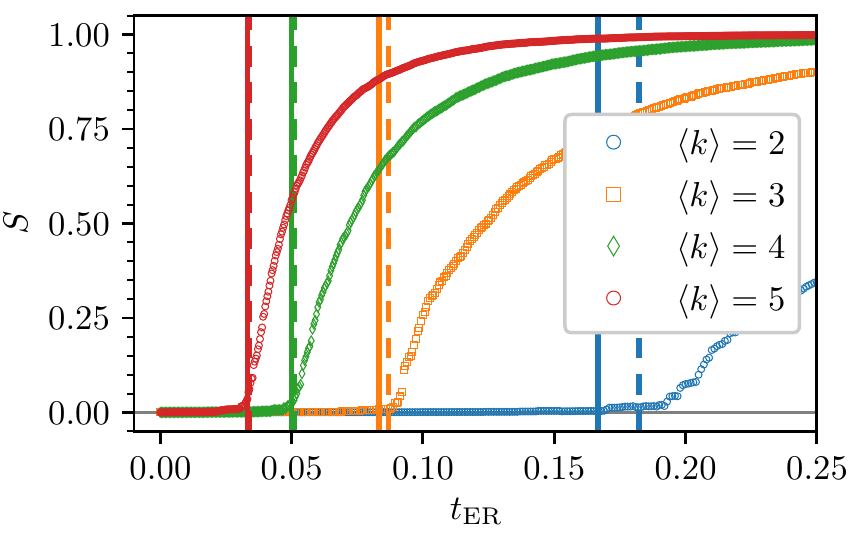}
  \caption{Proportion of nodes in the giant connected component $S$ as a function of time $t_{\text{ER}}$ for a few select mean degree $\enmean{k}$ networks. Solid vertical lines correspond to the critical time $t^{\text{c}}_{\text{ER}}$ while dashed vertical lines correspond to the revised critical time $t^{\text{r}}_{\text{ER}}$.}
  \label{fig:s_vs_t_er_zoom}
\end{figure}

The revised critical point in the ER case is thus
\begin{equation}
  t^{\text{r}} = -\log(1-t^{\text{c}}) = \log\left(\frac{\enmean{k}(\enmean{k}+1)}{\enmean{k}(\enmean{k}+1)-1}\right).
\end{equation}
\Cref{fig:s_vs_t_er_zoom} confirms that $t^\text{r}$ is a better predictor of the location of the phase transition. The difference between $t^{\text{c}}$ and $t^{\text{r}}$ becomes negligible as the mean degree increases.

Another aspect that could influence the position of the critical point is the possibility of having rewired more edges that needed to connect previously disconnected components. We show in the \cref{sec:extredges} that this should have no bearing on the critical point in large networks.

\section{\label{sec:conclusions}Discussion}
In this paper we studied the evolution of highly clustered networks under random edge rewiring dynamics. Our main result is showing the existence of a phase transition in which a giant connected component emerges. \citet{del_genio_endemic_2013} showed that equilibrium ensembles of degree-regular networks with prescribed clustering always admit a giant connected component. As a consequence, spreading processes such as infectious diseases in contact networks could always become endemic regardless of the level of clustering. By contrast, our work implies that this need not be the case in non-equilibrium systems. Depending on the precise mechanisms of time evolution of real networks and the level of clustering maintained, a giant connected component facilitating spreading processes may or may not exist. We have studied a model in which highly clustered populations undergo fully random connectivity changes and even in this simple scenario we observe two different modes of global connectivity.

Another interesting aspect of our work is from the perspective of statistical mechanics. A maximally clustered network is essentially the lowest entropy microstate in the context of the random network ensembles studied here. This is because such a network, under relabelling of nodes, is unique and least likely to be produced by chance at equilibrium. By using this configuration as a starting state for network dynamics, we have shown that the emergence of global connectivity is effectively delayed. This raises several other research questions, for example, is random rewiring the most or least effective method of delaying the onset of global connectivity? It is probable that more sophisticated rewiring methods involving choice, such as those studied in explosive percolation~\citep{souza2015}, would lead to different critical thresholds. We have also limited ourselves to studying rewiring that consistently destroys triangles, but what about rewiring with a view to increase the number of triangles? A number of greedy as well as equilibrium algorithms exist and are widely applied to model highly clustered networks~\citep{ritchie2004,foster_communities_2010}, but it is unclear how they cover the space of all networks and can lead to interesting behaviour such as hysteresis loops~\citep{foster_communities_2010}. Indeed, clustering in networks still leaves much to be explored.

\begin{acknowledgments}
J.K. was supported by the EPSRC under grant EP/IO1358X/1. S.J. is grateful for support from Spanish MINECO Grant No. FIS2013-43201-P (FEDER funds).
\end{acknowledgments}

\appendix*
\section{\label{sec:extredges}Extraneous edges}
Another mechanism that could change the location of the critical point $t^{\text{c}}$ is the number of extraneous edges between already connected components. A GCC is formed when there are enough external edges between the initial cliques. Only one external edge is needed to connect two cliques, but there are multiple ways to do it and sometimes multiple edges end up linking together the same cliques. For example, we need only one edge two join two disconnected triangles, but there are a total of 9 ways to do it, moreover there is no guarantee that we will not end up with multiple edges between these triangles.

More generally, let the average degree $\enmean{k}$ be fixed, then at $t=0$ there are $n=N/(\enmean{k}+1)$ cliques of size $\enmean{k}+1$. Any two cliques can therefore be connected in $(\enmean{k}+1)^2$ ways.

Suppose we never want to make more than one external edge to connect disconnected components. Then at $t=0$ the number of choices for placing an external edge is given by
\begin{equation}
  \left(\enmean{k}+1\right)\binom{n}{2} = \frac{N\left(N-\enmean{k}-1\right)}{2}.
\end{equation}
After each rewiring event, the number of choices decreases by $(\enmean{k}+1)^2$, so after $r-1$ rewires, the probability of placing an extraneous edge on the next rewire, $r$, is
\begin{equation}
  \mathbb{P}\left(\text{extra edge on step $r$}\right) = \frac{2\left(\enmean{k}+1\right)^2r}{N\left(N-\enmean{k}-1\right)}.
\end{equation}
Thus, the expected number of extraneous edges after $r$ rewiring events is
\begin{equation}
  \mathbb{E}\left(\text{extra edges by step $r$}\right) = \sum_{r^\prime=0}^{r}\frac{2\left(\enmean{k}+1\right)^2r^\prime}{N\left(N-\enmean{k}-1\right)}.
\end{equation}
In particular, setting $r=r^{\text{c}}=N/2(\enmean{k}+1)$ we get
\begin{multline}
  \mathbb{E}\left(\text{extra edges by step $r^{\text{c}}$}\right) \\ =\frac{2\left(\enmean{k}+1\right)^2}{N\left(N-\enmean{k}-1\right)}\cdot \frac{N}{4\left(\enmean{k}+1\right)}\left(\frac{N}{2\left(\enmean{k}+1\right)}+1\right) \\
  = \frac{N+2\enmean{k}+2}{4\left(N-\enmean{k}-1\right)}.
\end{multline}
Taking the limit $N\to\infty$, we get
\begin{equation}
  \mathbb{E}\left(\text{extra edges by step $r_c$}\right) \simeq \frac{1}{4},
\end{equation}
which is fixed and independent of network size. Therefore, the formation of extraneous edges does not affect the location of the critical point in the large network limit.

\bibliography{draft}

\end{document}